\documentclass[12pt]{article}
\usepackage{amssymb,amsmath}
\usepackage[noblocks]{authblk}
\usepackage[top=0.75in, bottom=0.75in, left=0.75in, right=0.75in, dvips]{geometry}
\usepackage{caption}
\usepackage{cite}
\usepackage{graphicx}
\usepackage{epstopdf}
\begin{document}
\textwidth 10.0in
\textheight 9.0in
\topmargin -0.60in
\title{An analytic approach for the study of pulsar spindown}
\author[1]{F.A. Chishtie}
\author[1]{S.R. Valluri}

\affil[1] {Department of Physics and Astronomy, The
University of Western Ontario, London, ON N6A 5B7, Canada}

\maketitle
\noindent


email: fchishti@uwo.ca, valluri@uwo.ca

\newcounter{saveeqn}
\newcommand{\alpheqn}{\setcounter{saveeqn}{\value{equation}}%
\stepcounter{saveeqn}\setcounter{equation}{0}%
\renewcommand{\theequation}{\mbox{\arabic{saveeqn}-\alph{equation}}}}
\newcommand{\reseteqn}{\setcounter{equation}{\value{saveeqn}}%
\renewcommand{\theequation}{\arabic{equation}}}

\newcounter{savesub}
\newcommand{\alphsub}{\setcounter{savesub}{\value{subsection}}
\stepcounter{savesub}\setcounter{subsection}{0}
\renewcommand{\thesubsection}{\mbox{\arabic{section}\hspace{2pt}\alph{subsection}}}}
\newcommand{\resetsub}{\setcounter{subsection}{\value{savesub}}
\renewcommand{\thesubsection}{\arabic{subsection}}}

\begin{abstract}
 In this work we develop an analytic approach to study pulsar spindown. We use the monopolar spindown model by Alvarez and Carrami\~{n}ana (2004), which assumes an inverse linear law of magnetic field decay of the pulsar, to extract an all-order formula for the spindown parameters which are expressed in terms of modified Bessel functions. We further extend the analytic model to incorporate the quadrupole term that accounts for the emission of gravitational radiation, and obtain expressions for the period $P$ and frequency $f$ in terms of transcendental equations. We derive the period of the pulsar evolution as an approximate first order solution in the small parameter present in the full solution.  We find that the first three spindown parameters of the Crab, PSR B1509-58, PSR B0540-69 and Vela pulsars are within their known bounds providing a consistency check on our approach. After the four detections of gravitational waves from binary black hole coalescence and a binary neutron merger 170814, which was a novel joint gravitational and electromagnetic detection, a detection of gravitational waves from pulsars will be the next landmark in the field of multi-messenger gravitational wave astronomy.   
\end{abstract}

\section{Introduction}

Pulsars are highly magnetized neutron stars which are known to rotate rapidly due to various physical mechanisms \cite{LK2005}. These mechanisms involve electromagnetic and gravitational emissions which result in the spindown of the GW frequency of the emitted pulsar signal \cite{AC2004}. 

In their paper on monopolar pulsar spindown, Alvarez and Carrami\~{n}ana have considered a general multipole spindown to study pulsar evolution \cite{AC2004}. In this model, a monopolar term was introduced to ensure that the braking indices defined in terms of the frequency and higher derivatives are in agreement with the trajectories in the $P\neg\dot{P}$ diagram for pulsar evolution, where $P$ and $\dot{P}$ denote the pulsar period and its time derivative. Their detailed analysis of the stationary multipole model ruled out the possibility of a time independent evolutionary equation for the pulsar frequency. Their time dependent multipole model included dynamics of the pulsar magnetic moment and thereby the decay of the magnetic field, $B(t)$. They show in their analysis that an inverse linear decay proposed by Chanmugham and Sang \cite{CG89}, in contrast to an exponential decay law for pulsar magnetic field \cite{Ostriker}, did a better fit of the evolutionary trajectories of the four pulsars studied namely the Crab, PSR B1509-58, PSR B0540-69 and Vela. Following their approach, we use an inverse linear decay law for $B(t)$. The spindown of pulsars due to the intense magnetic fields that surround them is a phenomenon that will significantly impact on the younger pulsars which can lose a large amount of rotational energy due to physical processes such as electromagnetic and gravitational multipole radiation. The evolution of pulsars has been studied in great detail \cite{Camilo, Chanmugam1, Chanmugam2, Colpi, ATNF, Lyne1, Lyne2, Manchester1, Manchester2, Manchester3, Manchester4, Manchester5, ES2011, Taylor, Ostriker, Alvarez3, Damour, Lyne2015}.

The first observations of gravitational waves by Advanced LIGO are identified as black hole mergers, namely GW150914 \cite{GW1}, GW151226 \cite{GW2} and a less significant candidate LVT151012 \cite{GW3}. Recently, GWs from another such merger, GW170104 has been reported \cite{GW4}. GW170814 was coherently observed by the advanced Virgo and two advanced LIGO detectors, produced by the coalescence of two stellar mass black holes. Recently, a neutron merger has been detected by both GW and electromagnetic observations \cite{GW5}. With increasing detector sensitivity of LIGO, VIRGO and the upcoming KAGRA, SKA and IndIGO detectors, the next wave of optimism and discovery should be on the detection of GW from pulsars. Pulsars are remarkably stable objects and although their GW amplitude is weaker compared to black holes, the fact that they can be tracked over a long period of time and that they emit continuous GW signals will be invaluable in detecting their GW signals.

We extend the mathematical analysis of the monopolar pulsar spindown model introduced by Alvarez and Carrami\~{n}ana \cite{AC2004, Alvarez1}. We further present a means of analytically finding the pulsar spindown parameters as defined by Brady et al. \cite{BC2000, BC1998} and Jaranowski et al. \cite{JK1999} for the GW frequency and the phase measured at the ground based detectors. 

Our paper is organized as follows. In the next section, we present a model for pulsar spindown along with a new solution that includes the quadrupole term due to gravitational radiation emission, which is an extension relevant for younger pulsars. In section 3, we connect this model to the form of frequency spindown shifts by Brady et al. \cite{BC2000}, via the orthogonal properties of the Chebyshev polynomials and determine an analytical form of the spindown parameters. We also compute the first three spindown parameters via this approach for the Crab, PSR 1509-58, PSR 0540-69 and Vela pulsars and find that these are within the stated bounds provided by \cite{BC2000}. These bounds provide a consistency check to the model. Section 4 summarizes our conclusions.

\section{A model for pulsar spindown}

We consider the spindown model derived from a general spindown law \cite{AC2004},
\begin{equation}
\dot{f}=-F(f,t)
\end{equation}

This law expresses the change of pulsar frequency with respect to time and assumes that the frequency is a positive, antisymmetric and continuous function of time. Following these properties, the even powers of $f$ are excluded and the first three non-zero terms in the Taylor expansion of Eq.(1) are, 

\begin{equation}
\dot{f}=-s(t)f-r(t)f^{3}-g(t)f^5
\end{equation}

The first term, namely the monopolar term was introduced by Alvarez and Carrami\~{n}ana (2004) to take into account particle acceleration mass loss or pulsar winds. This term further enables the braking indices to be in accord with the trajectories in the $P\neg\dot{P}$ diagrams for pulsar evolution. The second and third terms incorporate the normal spindown mechanisms of magnetic dipole radiation and gravitational radiation. For the frequencies measured in known, isolated pulsars, higher order terms can be neglected. In contrast, the spindown of millisecond pulsars has shown features strongly in contrast with typical pulsars. Hence, this model does not describe binary pulsars. A thorough analyses of binary pulsar spindown has been done in several excellent works \cite{Damour, Taylor}, to cite but a few.  

Since $f=\frac{1}{P}$ and $\dot{f}=\frac{-\dot{P}}{P^2}$

\begin{equation}
\dot{P}=s(t)P +\frac{r(t)}{P}+\frac{g(t)}{P^3}
\end{equation}

For a simpler analysis for older pulsars, the quadrupole term $g(t)$ is dropped, as it does not give an appreciable contribution:

\begin{equation}
\dot{P} \simeq s(t)P+\frac{r(t)}{P}
\end{equation}

The assumption is that the frequency/period evolution changes as the magnetic field $B(t)=B(0)/ \left(1+t/t_c \right)$ decays. Moreover, the $\{s(t),r(t)\}\propto B(0)^{2} \psi (\frac{t}{t_c})$; thereby $r_0$, $s_0$ are obtained from pulsar evolution studies\cite{AC2004}. \\

From the use of the inverse linear law in Eq.(4), we find,

\begin{equation}
\dot{P}=\left({\frac{1}{1+t/t_c}} \right)^{2}\left(\frac{r_0}{P}+s_{0}P \right)
\end{equation}

\noindent where $r_0 \geqslant 0, s_0 \geqslant 0$, and $t_c$ is the characteristic time of the magnetic field decay. \\

If we have the initial condition $P(t_0)=P_0$, then the solution of this equation is given below:

 \begin{subequations}
    \begin{align}
      P(t)=\frac{1}{s_0} \sqrt{s_{0} \exp \left \{\frac{2t_{c}^{2} s_0(t-t_0)}{(t+t_c)(t_c+t_0)} \right\} (P_{0}^{2}s_{0}+r_0)-r_0s_0} \\ \intertext{\indent For $s_0<0, r_0\ll |s_0| \ll1, t_c\gg t, t_0=0, g_0=0$, we have}
      P(t)=\sqrt{\frac{e^{2s_0t}\left(P_0^2s_0^2+r_0s_0\right)}{s_0^2}-\frac{r_0}{s_0}}\sim P_0e^{s_0t}
    \end{align}
 \end{subequations}

In the most general form, for $s_0>0, r_0>0$,

\begin{equation}
\begin{split}
f(t) &=s_0\left[\sqrt{s_0\exp\left(\frac{2t_c^2s_0(t-t_0)}{(t+t_c)(t_c+t_0)}\right)\left(\frac{1}{f_0^2}s_0+r_0\right)-r_0s_0}~\right]^{-1}\\
 &=f_0s_0^{1/2}\left[\sqrt{\lambda_0\exp\left(\frac{2t_c^2s_0(t-t_0)}{(t+t_c)(t_c+t_0)}\right)-\lambda_1}\right]^{-1}
\end{split}
\end{equation}
\noindent where $\lambda_0 = f_{0}^{2}r_0+s_0$, $\lambda_1 = r_{0}f_0^2$, and $f(0)=f_0$,

For $t_c\gg t $ and $t_0=0$,
 \begin{subequations}
    \begin{align}
    f(t)=\frac{f_{0}s_{0}^{1/2}}{\sqrt{\lambda_0\exp(2s_0t)-\lambda_1}} \\ \intertext{\indent It should be noted that Eq.(8a) can also be expressed to account for a possible negative sign in $s_0$ as}
     f(t)=\frac{f_0|s_0|}{\sqrt{s_0\lambda_0 \exp (2s_0t)-\lambda_1s_0}}
    \end{align}
 \end{subequations}

It should be observed that the inclusion of a quadratic term in $f$ for a case in which $f$ is not antisymmetric would not present undue problems in the solution of the nonlinear differential Eq.(2).

For pulsars such as PSR B1509-58, there is an initial phase of strong gravitational spindown with the quadrupole parameters being at least two orders of magnitude higher than that of the other pulsars. We extend this model to further include the quadrupole term. Inclusion of the quadrupole term to consider the effects of spindown to GW emission, gives

\begin{equation} 
\frac{dP}{dt}=\left(s_0P+\frac{r_0}{P}+\frac{g_0}{P^3}\right)\left(1+\frac{t}{t_c}\right)^{-2}
\end{equation}

Multiplying by $P^3$ we obtain

\begin{equation}
P^3dP=\left(s_0P^4+r_0P^2+g_0\right)\left(1+\frac{t}{t_c}\right)^{-2}dt
\end{equation}

\begin{equation}
\frac{P^3dP}{s_0P^4+r_0P^2+g_0}=\left(1+\frac{t}{t_c}\right)^{-2}dt
\end{equation}

Let $Q=P^2$, in the above differential equation. Thereby, $dQ=2PdP$, and

\begin{equation} \label{2}
\frac{QdQ}{2(s_0Q^2+r_0Q+g_0)}=\left(1+\frac{t}{t_c}\right)^{-2}dt
\end{equation}

\begin{equation}
\frac{QdQ}{2s_0(Q^2+\frac{r_0Q}{s_0}+\frac{g_0}{s_0})}=\frac{QdQ}{2s_0(Q+a)(Q+b)}
\end{equation}

Since
\begin{equation} \label{general}
\int \frac{QdQ}{(Q+a)(Q+b)}=\frac{1}{a-b} \{a \ln(Q+a)-b \ln(Q+b)\}
\end{equation}

Integrating both sides of Eq.(9), we obtain

\begin{equation} \label{3}
\frac{1}{a-b} \left \{a \ln(Q+a)-b \ln(Q+b) \right\}=\frac{-2t_cs_0}{1+t/t_c}+C
\end{equation}

The initial condition $P(t_0)=P_0$ can be used to fix $C$ , \\

which can be expressed in the form:
\begin{equation}
C=\frac{1}{a-b} \left\{a \ln(P_0^2+a)-b \ln(P_0^2+b) \right \}+\frac{2t_cs_0}{1+t_0/t_c}
\end{equation}

Therefore the full solution to Eq.(10) can be expressed as 
\begin{equation} \label{4}
\frac{1}{a-b} \left \{a \ln \left(\frac{P^2+a}{P_0^2+a}\right)-b \ln \left(\frac{P^2+b}{P_0^2+b}\right) \right\} =2t_cs_0 \left\{1-\frac{1}{1+t/t_c}\right\} \approx 2s_0t (t_c\gg t)
\end{equation}

Eq.(\ref{4}) is symmetric under interchange of $a$ and $b$, suggesting a very interesting pattern of the generalized Lambert W function \cite{AT2017}, which we explore in the next section. \\

We can solve for $a$ and $b$ by finding the roots of the quadratic expression of $Q^2+\frac{r_0}{s_0}Q+\frac{g_0}{s_0}$. Here $a+b=\frac{r_0}{s_0}$, $ab=\frac{g_0}{s_0}$, $a-b=\frac{\sqrt{r_0^2-4s_0g_0}}{s_0}$. \\

Hence, 
\addtocounter{equation}{0}
\begin{subequations}
\begin{align}
a=\frac{1}{2}\left (\frac{r_0+\sqrt{r_0^2-4s_0g_0}}{s_0}\right)\\
b=\frac{1}{2} \left(\frac{r_0-\sqrt{r_0^2-4s_0g_0}}{s_0}\right)
\end{align}
\end{subequations}

Here $s_0\geq 0$, $r_0 \geq 0$ and $g_0 \geq 0$ where $s_0 > r_0 > g_0$. By letting $b=a\epsilon $ for small $b$, we obtain 

\begin{equation*}
a \ln \left ( \frac{Q+a}{Q_0+a} \right ) + a\epsilon \ln \left ( \frac{Q_0+a\epsilon}{Q+a\epsilon} \right) = 2s_0ta-2s_0t a\epsilon
\end{equation*}
which simplifies to
\begin{equation} \label{13}
 \left [ \ln\left( \frac{Q+a}{Q_0+a}\right) - 2s_0t \right] + \epsilon \left [ \ln \left(\frac{Q_0+a\epsilon}{Q+a\epsilon}\right) +2s_0t \right] = 0
\end{equation}

It is worthwile to keep in consideration the possibility that the coefficients $s_0$, $r_0$,$g_0$ can change  sign to indicate the occurence of glitches in some time segments of pulsar data. Then there is a spinup in those time segments. Although Alvarez and Carrami\~{n}ana \cite{AC2004} considered the coefficients $s_0, r_0$ and $g_0$ to be $\geq 0$, to restrict the study to spindown of pulsars, it is worth to keep open the cases where one or more of  $s_0, r_0$ and $g_0$ could be $<0$, giving rise to the possibility that $a$ and/or $b$ could be $<0$. Such cases of negative values can indicate pulsar spinups, also known as glitches. Glitches are discrete changes in the pulsar rotation rate that is often followed by a relaxation \cite{ES2011, Lyne2015}. The cumulative effect of glitches is to reduce the regular long-term spindown rate $|\dot{f}|$ of the pulsar.

\subsection{The Lambert W Solution}
Under the special case of $a=b$, which may  be a rare situation when the expressions in the radical sign in Eqs.(18a) and (18b) vanish, the left hand side of Eq.(\ref{general}) can be integrated as

\begin{equation}
\int \frac{Q}{(Q+a)^2}dQ =\log (Q+a) + \frac{a}{Q+a} = -\frac{2t_cs_0}{1+ t/t_c} + C
\end{equation}

When $t=0$ and $Q=Q_0$,

\begin{equation}
\log \left (Q_0 + a \right) + \frac{a}{Q_0+a} =  C-2t_cs_0
\end{equation}

The solution can be concisely written as , 
\begin{equation}
-\log\left( \frac{a}{Q+a}\right) + \frac{a}{Q+a} + \log a = 2s_0t + \log \left ( Q_0+a \right ) + \frac{a}{Q_0+a}
\end{equation}

Substituting $\frac{a}{Q+a}=z$ and $\frac{a}{Q_0+a}=z_0$, we obtain by exponentiation on both sides

\begin{eqnarray}
e^{-\log z+z} = e^{2s_0t + z_0- \log z_0}\nonumber \\
\end{eqnarray}

Rearranging gives

\begin{equation}
-ze^{-z} = -z_0e^{-z_0}e^{2s_0t}
\end{equation}

This transcendental equation can be solved to yield

\begin{equation}
-z=W\left (-z_0e^{-z_0}e^{-2s_0t} \right)
\end{equation}

\noindent where $W$ is the Lambert W function \cite{Lambert}. Thus, again for $t_c \gg t$

\begin{equation}
-\frac{a}{Q+a} = W \left (-\frac{a}{Q_0+a} e^{-a/(Q_0+a)}  e^{-2s_0t} \right )
\end{equation}

\noindent where $Q$ and $Q_0$ as defined previously are $Q=P^2$ and $Q_0=P_0^2$.\\

$W(z)$ has the series expansion

\begin{equation}
\begin{split}
W(z)&=\sum_{n\geqslant 1}\frac{(-n)^{n-1}}{n!}z^n\\
&\approx z-z^2+\frac{3z^3}{2}+- - - -
\end{split}
\end{equation}

\subsection{Spindown as a function of frequency}

From Eq.(19), we give an equivalent expression in terms of the frequency
\begin{equation}
 \ln \left ( \frac{1}{f^2} +a \right ) - \epsilon \ln \left ( \frac{1}{f^2}+a\epsilon \right ) = (1-\epsilon) 2s_0t + \ln \left ( \frac{1}{f_0^2}+a \right ) - \epsilon \ln \left ( \frac{1}{f^2_0}+a\epsilon \right )
\end{equation}

\begin{equation}
\ln\left[\frac{\frac{1+af^2}{f^2}}{\left(\frac{1+a\epsilon f^2}{f^2}\right)^{\epsilon}}\right]=\ln\left[\frac{\frac{1+af_0^2}{f_0^2}}{\left(\frac{1+a\epsilon f_0^2}{f_0^2}\right)^{\epsilon}}\right]+ (1-\epsilon)2s_0t
\end{equation}

For $g_0 \ll r_0 < s_0$, $a$ given by Eq.(18a), with $b=a\epsilon$ ($\epsilon \approx 10^{-2}$) and $s_c=(1-\epsilon)s_0$, we obtain,

\begin{equation}
\left (\frac{1+af^2}{1+af_0^2} \right ) \left (\frac{f_0^2}{f^2} \right )= e^{2s_ct}
\end{equation}

which can be written as

\begin{equation}
\frac{1}{f^2}=\left(\frac{1+af_0^2}{f_0^2}\right)e^{2s_ct}-a
\end{equation}

For the Crab pulsar, for this case of using the inverse magnetic law, $r_0 = 7.5\times10^{-12}$ Hz$^{-1}$, $s_0=9.4 \times 10^{-15}$ Hz was determined in [2], however $g_0$ was not found as this parameter was assumed to be zero in their model.\\

There are three possible cases for the roots in Eqs.(18a) and (18b):
\begin{enumerate}
\item $r_0^2-4s_0g_0 > 0, a > b$, e.g. in the case of the Crab pulsar.
\item $r_0^2-4s_0g_0 = 0, a = b$, there are no data available for this case, but this case can arise from pulsars that have certain values for $s_0, r_0$ and $g_0$.
\item  $r_0^2-4s_0g_0 < 0$, $\sqrt{r_0^2-4s_0g_0}\pm i\sqrt{-r_0^2+4s_0g_0}$, where $a$ and $b$ are complex. The Vela is a fine example of where this could be possible.
\end{enumerate}

For $b=a\epsilon \ll  a$, $ g_0 \neq 0$, and $s_c = s_0(1-\epsilon)$ we approximately obtain, 

\begin{equation}
f(t)=\frac{1}{\left[\left(\frac{1+af_0^2}{f_0^2}\right)e^{2s_ct}-a\right]^{1/2}}
\end{equation}

Comparing the expression of $f$ here with that for $g_0=0$, 

\begin{equation}
f(t)=\frac{f_0s_0^{1/2}}{\sqrt{\lambda_0e^{2s_0t}-\lambda_1}}
\end{equation}

\noindent where $\lambda_0=f_0^2r_0+s_0, \lambda_1=r_0f_0^2, a=\frac{r_0}{s_0}$, we find that when $\epsilon=0$, the above two equations coincide exactly.\\

For $\epsilon \neq 0$,  Eq.(32) can be written as,
 
\begin{equation}
f(t) = \frac{f_0}{\left\{\left(1+af_0^2\right)e^{2s_ct}-af_0^2 \right\}^{1/2}}
\end{equation}

\begin{equation}
f(t)=\frac{f_0}{\left[(1+af_0^2)e^{2s_ct}\left\{1-\left(\frac{af_0^2}{1+af_0^2}\right)e^{-2s_ct}\right\}\right]^{1/2}}
\end{equation}
This can be written approximately, if one ignores the higher order terms in the curly brackets as
\begin{equation}
f(t) =f_0\left(1+af_0^2\right)^{-1/2}e^{-s_ct}
\end{equation}

It should be noted that for $g_0\neq 0$, 
\begin{equation}
1+af_0^2=1+\frac{r_0}{s_0}f_0^2+Cf_0^2
\end{equation}

\noindent where $C=\frac{1}{2} \left(-\frac{2s_0g_0}{r_0^2}+\ldots\right)$ in the expansion of the radical $\sqrt{1-\frac{4s_0g_0}{r_0^2}}$. Consequently, $\lambda_0$ will not be the same as given previously ($\lambda_0=s_0+r_0f_0^2$), but will be

\begin{equation}
\lambda_c=\lambda_0+s_0Cf_0^2+\ldots
\end{equation}

Hence, 

\begin{equation}
f_{pulsar}=f_0\sqrt{\frac{s_0}{\lambda_c}}e^{-s_ct}\left[1-\frac{\lambda_1}{\lambda_c}e^{-2s_ct}\right]^{-1/2}
\end{equation}

For $g_0 \neq 0$, simplification of Eq.(17a) in terms of $\lambda_0=r_0f_0^2+s_0$ gives

\begin{equation}
a=\frac{r_0}{2\left(\lambda_0-f_0^2r_0\right)}\left\{1+\frac{|r_0|}{r_0}\sqrt{1-4g\left(\frac{\lambda_0}{r_0^2}-\frac{f_0^2}{r_0}\right)}\right\}
\end{equation}

\noindent where the $|r_0|$ addresses the situation where $r_0<0$.

This expression can be more concisely written as
\begin{equation}
a=\frac{r_0}{2\Lambda}\left\{1+\frac{|r_0|}{r_0}\sqrt{1-\frac{4g}{r_0^2}\Lambda}\right\}
\end{equation}

\noindent where $\Lambda=\lambda_0-f_0^2r_0$.

\section{Gravitational Wave Signal with Spindown Corrections}

Based on the derivations done in section 2, we have two cases: $g_0=0 \ (\epsilon=0)$ and $g_0\neq 0 \ (\epsilon \neq 0)$. We explore both cases below. 

It should be noted that time-dependent $r_0, s_0$ are taken from Table 3. in Alvarez and Carrami\~{n}ana \cite{AC2004}, assuming an inverse linear magnetic field decay timescale consistent with $r_0\geq0$ and $s_0\geq 0$. $g_0$ is taken from the stationary multipole model in Table 1.\cite{AC2004}, which also contains values of $r_0$ and $s_0$, which differ little from those in Table 3. \cite{AC2004}. Better values of $r_0$, $s_0$ including the presently undetermined $g_0$ should be available from pulsar data in the coming years.

\subsection*{a. $g_0=0 ~ (\epsilon=0)$}

\begin{equation}
f_{pulsar}(t)=f_0 \sqrt{\frac{s_0}{\lambda_0}} \exp \left[\frac{-s_{0}t}{1+t/t_c}\right]{\left[1-\frac{\lambda_1}{\lambda_0} \exp \left\{\frac{-2s_{0}t}{1+t/t_c}\right\}\right]}^{-1/2}
\end{equation}

Using the binomial expansion we can rewrite this as $(1-x)^{-1/2}=1+\frac{1}{2}x+\frac{3}{8}x^2+...$, thus

\begin{equation}
f_{pulsar}(t)=f_{0}\sqrt{\frac{s_0}{\lambda_0}} \exp \left\{\frac{-s_0t}{1+t/t_c}\right\} \left[1+\frac{\lambda_1}{2\lambda_0} \exp \left\{\frac{-2s_0t}{1+t/t_c}\right\} + \frac{3\lambda_1^2}{8x_0^2} \exp \left\{\frac{-4s_0t}{1+t/t_c}\right\} + ...\right]
\end{equation}

Higher order terms were dropped. Since $t/t_c \ll 1$, we have 

\begin{equation}
f_{pulsar}(t)=f_0\sqrt{\frac{s_0}{\lambda_0}} \exp \{-s_0t\} \left[1+\frac{\lambda_1}{2\lambda_0} \exp \{-2s_0t\} + ...\right]
\end{equation}

The spindown of GW signal from pulsars has been studied in pioneering works by a parametrized model for the gravitational wave frequency \cite{BC2000, JK1999}:

\begin{equation}
f=f_0\left(1+\frac{\vec{v}}{c} \cdot \hat{n}\right)\left(1+\sum_{k=1} f_k\left[t+\frac{\vec{x}}{c}\cdot \hat{n}\right]^k \right)
\end{equation}
\noindent where the terms $\frac{\vec{v}}{c} \cdot \hat{n}$ and $\frac{\vec{x}}{c}\cdot \hat{n}$ account for the Doppler shift. By ignoring the Doppler shift (which we investigate in a later paper), we can further express the GW signal in terms of the parametrized series for pulsars as a linear combination of Chebyshev polynomials:

\begin{equation}
f_{GW}(t)=\sum_{k} f_k\left(\frac{t}{\tau_{min}}\right)^k = \sum_{k} f_k T_k \left(\frac{t}{\tau_{min}}\right)
\end{equation}

Let $x = t/ \tau_{min}$, where $T_k(x)$ are the Chebyshev polynomials of the first kind.

Therefore,

\begin{equation}
f_{GW}(t)=\sum_{k} f_k T_k(x)
\end{equation}

\noindent and $f_k$ are the spindown parameters. \\

We can evaluate the spin-down parameter by using the orthogonal properties of the Chebyshev polynomials, by using the equation

\begin{equation}
f_{GW}(t) \approx f_{pulsar}(t)
\end{equation}

Multiplying by $\frac{T_l(x)}{\sqrt{1-x^2}}$ on both sides, we obtain

\begin{equation}
\frac{T_l(x)f_{GW}(t)}{\sqrt{1-x^2}} = \frac{T_l f_{pulsar}(t)}{\sqrt{1-x^2}}
\end{equation}

Integration over the domain $[-1,1]$ gives, 

\begin{equation}
\int_{-1}^{1} \sum_{k} f_k \frac{T_l(x)T_k(x)}{\sqrt{1-x^2}} dt = \int^{1}_{-1} \frac{T_l f_{pulsar}(t)}{\sqrt{1-x^2}} dt.
\end{equation}

From the orthogonality conditions of Chebyshev polynomials, 

\begin{align}
\int_{-1}^{1} \frac{T_j(x)T_k(x)}{\sqrt{1-x^2}} dx = \left\{ \begin{array}{cc} 
                \pi & \hspace{5mm} j=k=0 \\
                \frac{\pi}{2} & \hspace{5mm} j=k\neq 0 \\
                 0 & \hspace{5mm} j\neq k\\
                \end{array} \right.
\end{align}

\noindent we obtain the following expression for the spindown parameters,

\begin{equation}
f_k=\frac{2}{\pi} \int^{1}_{-1} \frac{T_k(x)f_{pulsar}(t)}{\sqrt{1-z^2}} dt
\end{equation}

Evaluating the integral enables us to find $f_k$,

\begin{equation}
\int_{-1}^{1} \frac{1}{\sqrt{1-z^2}} e^{-pz} T_k(z) dz = (-1)^{k} I_k(p)
\end{equation}

\noindent where $I_k(p)$ is the modified Bessel function. \\

For the Crab pulsar with $\tau_{min}\approx 962$ years, we have

\begin{equation}
\frac{1}{\tau_{min}}\approx 3.296 \times 10^{-11} s^{-1}
\end{equation}

\begin{equation}
f_1 \approx -5.1804 \times 10^{-12} s^{-1} < \frac{1}{\tau_{min}}
\end{equation}

\begin{equation}
f_2 \approx 9.6924 \times 10^{-24} s^{-2} < \frac{1}{\tau^2_{min}}=1.0865 \times 10^{-21} s^{-2}
\end{equation}

\begin{equation}
f_3 \approx -1.2103 \times 10^{-35} s^{-3}  < \frac{1}{\tau^3_{min}} = 3.581 \times 10^{-32} s^{-3}
\end{equation}

The period of the Crab pulsar is $33.5 \times 10^{-3} s$, hence the frequency is $29.8508~Hz$. \\

\begin{table}
\centering
\begin{tabular}{|c | c | c | c| c| c|} 
 \hline
 Pulsar & Age (years) & $f_0$ (Hz) & $f_1 (s^{-2})$ & $f_2 (s^{-3})$ & $f_3 (s^{-4})$ \\ [0.5ex] 
 \hline\hline
 Crab & 962 & 29.937 &  $-5.180 \times 10^{-12}$ & $9.692 \times 10^{-24}$ & $-1.210 \times 10^{-35}$ \\ 
\hline
 PSR1509-58 & 1553 & 6.627 & $-1.657 \times 10^{-12}$ & $1.988\times 10^{-24}$ & $-1.591\times 10^{-36}$\\
\hline
 PSR0540-69	& 1664 & 19.881 & $-4.405\times 10^{-12}$ & $1.101\times 10^{-23}$ & $-1.835\times 10^{-35}$\\
\hline
 Vela & 11000 &11.198 & $-8.499 \times 10^{-13}$ & $3.612 \times 10^{-25}$ & $ -1.023 \times 10^{-37}$\\
 \hline
\end{tabular}
\caption{First three spindown parameters for the Crab, PSR1509-58, PSR0540-69 and Vela pulsars}
\label{table:1}
\end{table}

The values of the calculated spindown parameters are shown in Table 1 above and within the prescribed limit $|{f_k}| \leq \frac{1}{\tau^k_{min}}$, which is consistent with the parameterization of Brady \& Creighton and Jaranowski et al.\cite{BC2000, JK1999}. This is also a good indication of the robustness of the model by Alvarez and Carrami\~{n}ana \cite{AC2004} where $r_0$ and $s_0$ were determined independently of $f_1, f_2$ and $f_3$.

\subsection*{b. $g_0\neq 0~(\epsilon \neq 0)$}

 $f_{pulsar}(t)$ now assumes the form

\begin{equation}
f_{pulsar}(t) \approx f_0 \sqrt{\frac{s_0}{\lambda_c}} e^{-s_ct}
\end{equation}

\noindent where higher order terms are dropped, but can be included if a more accurate approximation is warranted. \\

Hence, 

\begin{equation}
f_k=\ \sqrt{\frac{s_0}{\lambda_c}} (-1)^{k} I_k (s_c)
\end{equation}

For $s_0>0, g_0\neq 0, \lambda_c <\lambda_0$, the spindown coefficient $|f_k|$ will be having slightly higher values in comparison to the case when $g_0=0$. Also when $g_0<0$, as can occur for spinups, $C$ will be positive and $|f_k|$ will be lower during such time segments. For $s_0<0$, the expression for $f(t)$ is modified accordingly as demonstrated in the limiting case of $g_0 = 0$ in Eq.(8b) above, whereby higher values of the spindown parameters are obtained. \\

For a relatively old pulsar, such as the Vela pulsar, only three spindown parameters may be adequate. For a young pulsar, $g_0$ could be more significant and more spindown parameters need to be evaluated. Eq.(52) provides the analytic expression for $f_k$, which could be used for all spindown parameters.

\section{Conclusions}

In earlier works \cite{JVD96, CQG2002, VCV2005, CVRSW, NO2008} we have implemented the
Fourier transform (FT) of the Doppler shifted GW signal from a
pulsar with the Plane Wave Expansion in Spherical Harmonics
(PWESH). It turns out that the consequent analysis of the Fourier
Transform (FT) of the gravitational wave (GW) signal from a pulsar has a very
interesting and convenient development in terms of the resulting
spherical Bessel, generalized hypergeometric, Gamma and Legendre
functions.  These works considered frequency modulation of a GW
signal due to rotational and circular orbital motions of the
detector on the Earth.  In later analysis, rotational and
orbital eccentric motions of the Earth, as well as perturbations
due to Jupiter and the Moon were considered \cite{VCV2005}. The numerical analysis of this
analytic expression for the signal offers a challenge for
efficient and fast numerical and parallel computation.

The recent detection of gravitational waves from black hole mergers is an outstanding success of theoretical physics and experimental
gravitation. Gravitational wave detectors like the LIGO,
VIRGO, LISA, KAGRA and GEO 600 are opening a new window
for the study of a great variety of nonlinear curvature phenomena.
Detection of GW necessitates sufficiently long observation periods
to attain an adequate Signal to Noise ratio (S/N). The data analysis for
continuous GW, for example from rapidly spinning neutron stars, is
an important problem for ground based detectors that demands
analytic, computational and experimental ingenuity. The Crab and Vela pulsars are among the iconic sources of GW emissions. Abbott et al. have presented direct upper limits on GW emissions from the Crab pulsar \cite{Abbott2008}. The searches use the known frequency and position of the Crab pulsar. They find that, under the assumption that GW and the electromagnetic signals are phase locked, their single template search results constrain the GW luminosity to be less than 6\% of the observed spindown luminosity, and beats the indirect limits obtained from all electromagnetic observations of the Crab pulsar and nebula. Similarly, Abadie et al. have given the direct upper limits on GW emissions from the Vela pulsar using data from the VIRGO detector's second science run \cite{Abadie}.
The Square Kilometer Array (SKA) will soon be in operation, and along with Advanced LIGO, VIRGO, KAGRA and the upcoming IndIGO, direct detection of GW from pulsars may become a reality and serve as a landmark in the field of multi-messenger grativational wave astronomy in the very near future.

In this work, we have presented an analytic formulation 
for determining spindown parameters in the Brady and Creighton approach \cite{BC2000} using a pulsar model which assumes an inverse linear law decay of the magnetic field \cite{AC2004}. We were able to extract these parameters using the exact solution involving the monopolar, dipolar and quadrupolar terms in the model and found these to be proportional to the modified Bessel functions. For the Crab, PSR 1509-58, PSR 0540-69 and Vela pulsars, we obtained the first three spindown parameters which were within the limit $|{f_k}| \leq \frac{1}{\tau^k_{min}}$. This is consistent with the parameterization of Brady \& Creighton \cite{BC2000} and Jaranowski et al. \cite{JK1999}, and a good indication of the robustness of the model by Alvarez and Carrami\~{n}ana \cite{AC2004}. Further, we were able to find the full solution for the period (frequency) evolution. With the determination of the quadrupole coefficient $g_0$ from data, our solution can be incorporated for further improvement in accuracy of the spindown parameters.The study of pulsar spindown and evolution of its braking index will lead to further interesting explorations of the anomalies present in the timing structure, not only in connection with gravitational waves, but also in the fundamental aspects of quark deconfinement in pulsar cores \cite{Weber}. The study of pulsar spindown also implicitly involves the role of spinups. The physics behind glitches is an active ongoing area of research that presents challenging studies such as the interior of neutron stars and the properties of matter at ultra high nuclear densities \cite{Baym}. We are presently working on utilizing the available data on isolated pulsars towards finding fits for the $P\neg\dot{P}$ diagram that would also include the quadrupole term $g_0$ using the analytic expression derived in this work to study gravitational wave data mining for the Crab Pulsar \cite{FVK2}. We hope to further improve the accuracy of the spindown parameters for GW signal detection and extend the applicability of our approach to younger pulsars. In forthcoming work, we plan to develop the analytic Fourier Transform of the pulsar GW signal to include spindown \cite{FV1}.  

\section{Acknowledgments}

Firstly, we thank the anonymous reviewers of our earlier paper in CQG (2006) for their thorough and inspiring critique that has stimulated our continued efforts to tackle this problem. SRV would like to thank the Natural Sciences Engineering Research Council (NSERC) Canada for funding support through an NSERC Discovery Grant, and King's University College for their continued support in our research endeavors. We would like to also thank Vladimir Dergachev, Ken Roberts, Xingyi Wang and Justin Tonner for useful discussions.

\clearpage
\centering

\end{document}